\documentclass[%
prl,
twocolumn,
groupedaddress,
showpacs,preprintnumbers,
amsmath, amssymb, aps,raggedbottom
]{revtex4-1}
\usepackage{color}
\usepackage{amsmath}
\usepackage{amssymb}
\usepackage{graphicx}% Include figure files
\usepackage{epsfig}
\usepackage{epstopdf}
\usepackage{dcolumn}% Align table columns on decimal point
\usepackage{bm}% bold math
\usepackage{natbib}
\usepackage{textcomp}

\begin{document}

\title{Supersymmetric Optical Structures}% Force line 

\author{Mohammad-Ali Miri$^{1}$}\email{miri@knights.ucf.edu} \author{Matthias Heinrich$^{1}$} \author{Ramy El-Ganainy$^{2}$} \author{Demetrios N. Christodoulides$^{1}$}
\affiliation{$^{1}$CREOL$/$College of Optics, University of Central Florida, Orlando, Florida, USA\\$^{2}$Department of Physics, University of Toronto, 60 St. George Street, Toronto, Ontario, Canada M5S 1A7}
%\date{\today}
\begin{abstract}
%\noindent
We show that supersymmetry can provide a versatile platform in synthesizing a new class of optical structures with desired properties and functionalities. By exploiting the intimate relationship between superpatners, one can systematically construct index potentials capable of exhibiting the same scattering and guided wave characteristics. In particular, in the Helmholtz regime, we demonstrate that one-dimensional supersymmetric pairs display identical reflectivities and transmittivities for any angle of incidence. Optical SUSY is then extended to two-dimensional systems where a link between specific azimuthal mode subsets is established. Finally we explore supersymmetric photonic lattices where discreteness can be utilized to design lossless integrated mode filtering arrangements.
\end{abstract}

% PACS, the Physics and Astronomy

\maketitle
Supersymmetry (SUSY) emerged within quantum field theory as means to relate fermions and bosons \cite{Ram1971,Nev1971,Gel1971,Vol1973,Wes1974,Wit1981}. In this mathematical framework, these seemingly very different entities constitute superpartners and can be treated on equal footing. Transitions between their respective states require transformations between commuting and anti-commuting coordinates--better known as supersymmetries. The development of SUSY was also meant to resolve questions left unanswered by the standard model \cite{Bin2006}, such as the origin of mass scales or the nature of vacuum energy, and to ultimately link quantum field theory with cosmology towards a Grand Unified Theory. Moreover, SUSY has found numerous applications in random matrix theory and disordered systems \cite{Haake,*Efetov,*ZinnJustin}. Even though the experimental validation of SUSY is still an ongoing issue, some of its fundamental concepts have been successfully adapted to non-relativistic quantum mechanics (QM). Interestingly, in this context, SUSY has led to new methods in relating Hamiltonians with similar spectra. In this regard, it has been used to identify new families of analytically solvable potentials and to enable powerful approximation schemes \cite{Coo1995,Lah1990,Coo1983,Ste1985_1,*Ste1985_2}. Recently, SUSY schemes have been theoretically explored in quantum cascade lasers \cite{Bai2006} and ion-trap arrangements \cite{Yu2010}. Clearly of interest will be to identify other physical settings where the rich structure of SUSY can be directly observed and fruitfully utilized.

In quantum mechanics, SUSY establishes a relationship between superpartners through the factorization of an operator, i.e., $\mathcal{L}^{(1)}=A^{\dagger}A$, where $\dagger$ denotes the Hermitian adjoint. In this respect, the superpartner is defined through $\mathcal{L}^{(2)}=AA^{\dagger}$, from where one finds that $A\mathcal{L}^{(1)}=AA^{\dagger}A=\mathcal{L}^{(2)}A$ and $A^{\dagger}\mathcal{L}^{(2)}=A^{\dagger}AA^{\dagger}=\mathcal{L}^{(1)}A^{\dagger}$. It then follows that the two eigenvalue problems $\mathcal{L}^{(1,2)}X^{(1,2)}=\Omega^{(1,2)}X^{(1,2)}$ yield identical spectra $\Omega^{(1)}=\Omega^{(2)}$. Moreover, the SUSY operators $A^{\dagger}$ and $A$ pairwise transform the eigenfunctions of the respective potentials into one another: $X^{(1)}\propto A^{\dagger}X^{(2)}$ and $X^{(2)}\propto AX^{(1)}$ \cite{Coo1995}. In addition, supersymmetry demands that $A$ annihilates the ground state of $\mathcal{L}^{(1)}$. Therefore the corresponding eigenvalue is removed from the spectrum of $\mathcal{L}^{(2)}$. If however $A$ does not annihilate the ground state of $\mathcal{L}^{(1)}$, then the two operators share the exact same spectrum (including the fundamental state), and SUSY is said to be broken. In the language of superpotentials, this may also be characterized through the Witten parameter \cite{Wit1981,Coo1995}.

In this Letter we show that optics can provide a fertile ground where the ramifications of SUSY can be explored and utilized to realize a new class of functional structures with desired characteristics. In particular we demonstrate that supersymmetry can establish perfect phase matching conditions between a great number of modes-an outstanding problem in optics. In this vein, we illustrate the intriguing possibility for preferential mode-filtering where the fundamental mode of a structure can be selectively extracted. Moreover, in the Helmholtz regime, SUSY endows two very different scatterers with identical reflectivities and transmittivities irrespective of the angle of incidence. Subsequently we extend the concept of optical SUSY to two-dimensional (2D) settings with cylindrical symmetry, as in optical fibers. We show that a partner potential with a SUSY spectrum of radial modes exists, offering the possibility for angular momentum multiplexing. Finally, we investigate the implications of supersymmetry within the framework of finite periodic structures and propose a versatile approach to systematically design SUSY optical lattices.

To explore the consequences of supersymmetry in optics, we consider optical wave propagation in an arbitrary one-dimensional refractive index distribution $n(x)$. Waves propagating in the $xz$-plane can always be decomposed in their transverse electric (TE) and transverse magnetic (TM) components. For TE waves the field evolution is governed by the Helmholtz equation $\left ({\partial}_{xx}+{\partial}_{zz}+k_0^2n^2(x)\right )E_y (x,z)=0$. Modes propagating in this system have the form $E_y(x,z)=f(x)e^{i\beta z}$ and satisfy the following eigenvalue equation for the propagation constant $\beta$:
\begin{equation}
\label{eq1}
\mathcal{H}f(x)=-{\beta}^2f(x),
\end{equation}
\begin{figure}[b]
\begin{center}
\includegraphics[width=1\linewidth]{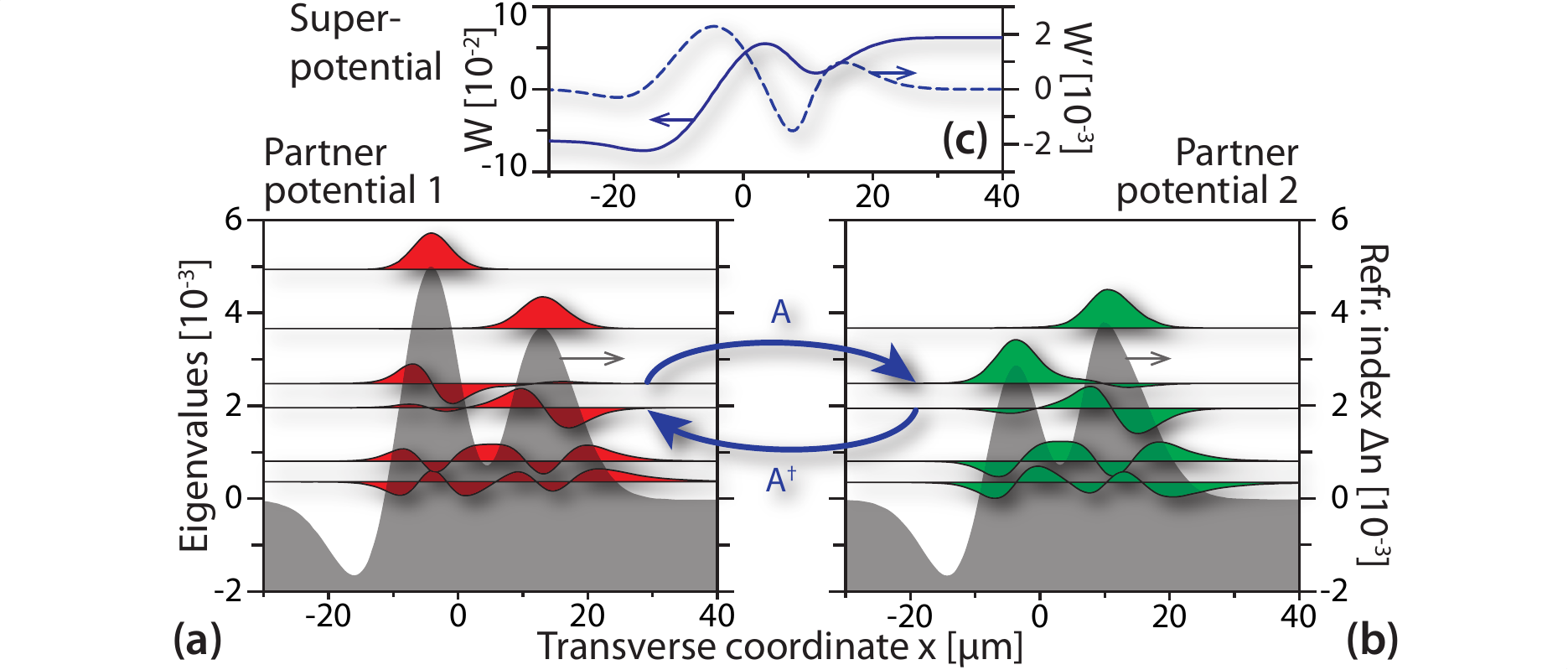}
\caption{(Color online). (a) Exemplary refractive index landscape (grey area) and its six bound modes (vertical placement indicates their respective eigenvalues). (b) SUSY partner and its five modes. The operators $A,A^{\dagger}$ transform the phase-matched  modes into each other. (c) Both index landscapes can be constructed from the superpotential $W$ and its slope $W'$.}
\label{fig1}
\end{center}
\end{figure}
where $\mathcal{H}=-\frac{d^2}{dx^2}-k_0^2n^2(x)$ corresponds to the Hamiltonian operator in a Schr\"odinger equation. For a given index profile $n^{(1)}(x)$, SUSY now provides a systematic way for generating a superpartner $n^{(2)}(x)$. If the index distribution $n^{(1)}(x)$ supports at least one bound state $f_1^{(1)}(x)$ (the ground state) with a propagation eigenvalue ${\beta}_1^{(1)}$, SUSY can be established via $\mathcal{H}^{(1)}+{\left({\beta}_1^{(1)}\right)}^2=A^{\dagger}A$, where $A=+d/dx+W(x)$ and $A^{\dagger}=-d/dx+W(x)$ are defined in terms of a yet to be determined superpotential $W(x)$. The optical potential and its superpartner then satisfy
\begin{equation}
\label{eq2}
{\left(k_0n^{(1,2)}(x)\right)}^2={\left({\beta}_1^{(1)}\right)}^2-W^2\pm W'.
\end{equation}
Taking into account that $A^{\dagger}Af_1^{(1)}=0$, one finds that $Af_1^{(1)}=0$. Thus a valid solution for $W$ can be obtained from the logarithmic derivative of the node-free fundamental mode:
\begin{equation}
\label{eq3}
W(x)=-\frac{d}{dx}\ln \left (f_1^{(1)}(x)\right ).\\
\end{equation}
\begin{figure}[t]
\begin{center}
\includegraphics[width=1\linewidth]{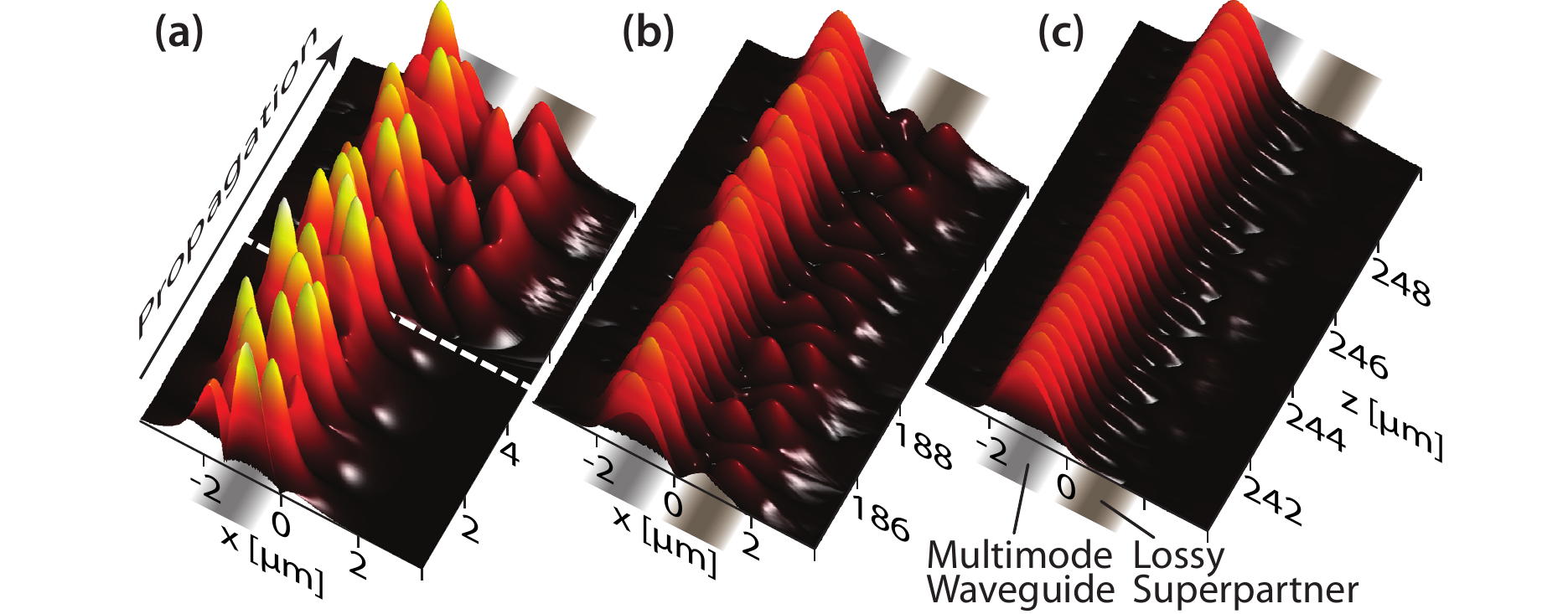}
\caption{(Color online). Beam propagation in a multimode waveguide. (a) When isolated (before dashed line), and when coupled to its lossy superpartner (after dashed line, losses: $\alpha\approx 0.4\textrm{cm}^{-1}$). Two more advanced stages of this same field evolution in the coupled system are shown in (b,c).}
\label{fig2}
\end{center}
\end{figure}
Figure~\ref{fig1}(a) depicts an arbitrary refractive index distribution supporting a set of six guided modes. Here the maximum index contrast is $5\times 10^{-3}$ and the wavelength used is $1\textrm{\textmu m}$. While Eqs.~(1-3) are valid in the Helmholtz regime, here we consider a low contrast structure that is experimentally feasible. For this example, the SUSY partner (Fig.~\ref{fig1}(b)) has been numerically calculated from Eq.~\eqref{eq2} through the corresponding superpotential (Fig.~\ref{fig1}(c)) that was obtained from Eq.~\eqref{eq3}. As Fig.~\ref{fig1}(a) clearly shows, the fundamental mode of $n^{(1)}$ lacks a partner in the eigenvalue spectrum of $n^{(2)}$, indicating unbroken SUSY. On the other hand the second state of $n^{(1)}$ is paired with the first mode of $n^{(2)}$ that has exactly the same propagation constant in spite of its different parity. In this way, all the modes of these two superpartners can be perfectly phase-matched except for the fundamental mode of $n^{(1)}$. Therefore SUSY provides the only strategy we know of to achieve global phase matching conditions, irrespective of how large the number of modes is, in such multimode optical potentials.\\ 
This latter feature can be exploited for mode filtering applications. The idea is illustrated in Fig.~\ref{fig2}(a) where $n^{(1)}$ has the form of a step-index like waveguide that supports three modes at $\lambda=1.5\textrm{\textmu m}$. The optical propagation when this system is excited with an arbitrary input beam, is depicted in the first propagation section of this figure. In this range, the field evolution is seemingly chaotic because of modal interference. Once however the superpartner waveguide is put in proximity, all the modes of $n^{(1)}$ (apart from the fundamental) are periodically coupled between these two structures. Despite their parity, coupling between the phase-matched modes occurs through their evanescent tails. If for example the second waveguide is made intentionally lossy, all the modes of $n^{(1)}$ eventually disappear except the fundamental, as shown in Figs.~\ref{fig2}(b,c). Similarly, the fundamental mode can be selectively amplified. This behavior could be potentially useful in large mode area laser sources.\\
\begin{figure}[t]
\begin{center}
\includegraphics[width=1\linewidth]{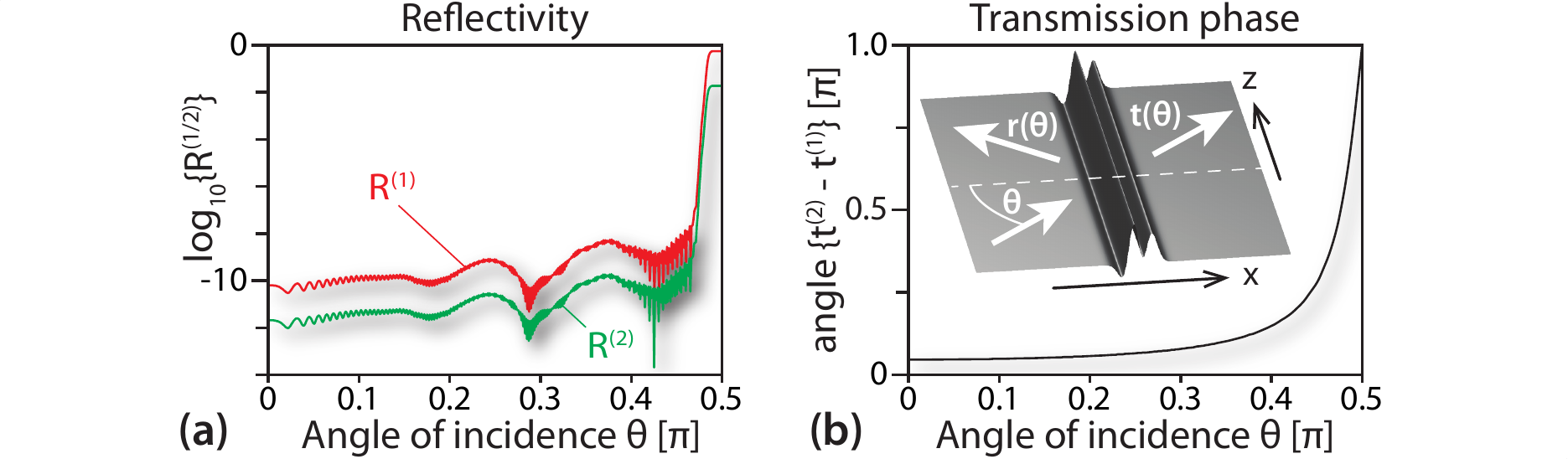}
\caption{(Color online). Scattering properties of the SUSY pair from Fig.~\ref{fig1}; (a) logarithmic plot of the angle-dependent reflectivities $R^{(1,2)}$ (the graphs have been offset for visibility), and (b) Phase difference of the transmission coefficients $t^{(1,2)}$ (inset: Schematic of the scattering configuration)}
\label{fig3}
\end{center}
\end{figure}
SUSY structures also exhibit identical scattering properties in terms of their reflectivities and transmittivities. In this case, the radiation mode continua are related to each other through the SUSY algebra. Let us consider again the SUSY pair described by Eqs.~\eqref{eq2}. We also assume that $n^{(1)}$ asymptotically approaches a constant background value $n_{\infty}$ at $x\rightarrow \pm\infty$. For an angle of incidence $\theta$, the components of the incident wave vector are $k_x=k_0n_{\infty}\cos(\theta)$ and $k_z=k_0n_\infty\sin(\theta)$. The SUSY formalism then relates the field reflection/transmission coefficients $r^{(1,2)}$ and $t^{(1,2)}$ associated with these two structures in the following way \cite{suppl}:
\begin{equation}
\label{eq4}
 r^{(2)}=+\frac{W_{\infty}+ik_x}{W_{\infty}-ik_x} r^{(1)},\quad
 t^{(2)}=-\frac{W_{\infty}+ik_x}{W_{\infty}-ik_x} t^{(1)},
\end{equation}
where $W_{\infty}=\sqrt{{\left({\beta}_1^{(1)}\right)}^2-k_0^2n_{\infty}^2}$ represents the limit of the superpotential at $x\rightarrow +\infty$ as obtained from Eqs.~\eqref{eq2}. Note that the argument of the square root is always a non-negative quantity \cite{Sny1983}. It follows that $n^{(1,2)}$ exhibit identical reflectivities $R^{(1)}=R^{(2)}={\left |r^{(1,2)}\right |}^2$ and transmittivities $T^{(1)}=T^{(2)}={\left |t^{(1,2)}\right |}^2$. Consequently, barring direct phase measurements, the two SUSY structures would be indistinguishable at any angle of incidence. Interestingly, the phase difference between $r^{(1)}$ and $r^{(2)}$, and between $t^{(1)}$ and $t^{(2)}$ for any given $\theta$ is solely determined by the propagation constant ${\beta}_1^{(1)}$ of the fundamental mode and the background refractive index $n_{\infty}$.\\
A schematic of a possible scattering arrangement is depicted in Fig.~\ref{fig3}. The angle-dependent reflection/transmission coefficients for the SUSY pair considered in Fig.~\ref{fig1}(a,b) were evaluated by means of the differential transfer matrix method \cite{Yeh1988} when the background refractive index is $n_{\infty}=1.5$. In accordance with our previous discussion, the two structures display identical reflectivities (Fig.~\ref{fig3}(a)). The phase difference between their respective transmission coefficients is also shown in Fig.~\ref{fig3}(b).\\
Having investigated SUSY in 1D optical systems, the question naturally arises as to whether these concepts can be extended to 2D structures. The answer is not particularly obvious given that the aforementioned factorization technique relies on 1D Hamiltonians \cite{Coo1995}. In what follows, we show that this limitation can be overcome in paraxial settings with cylindrical symmetry, as in weakly guiding optical fibers. 
In this regard, let us consider the radial refractive index profile $n(r)=n_\infty+\Delta n(r)$ where $\Delta n\ll n_\infty$. In this case, the slowly varying field envelope $U$ satisfies the paraxial equation
\begin{equation}
\label{eq5}
\left (-\frac{\partial^2}{\partial \eta^2}-\frac{1}{\eta}\frac{\partial}{\partial \eta}-\frac{1}{\eta^2}\frac{\partial^2}{\partial \phi^2}-V(\eta)\right )U=i\frac{\partial}{\partial \xi}U,
\end{equation}
where $\eta=r/r_0$ is a normalized radial coordinate, $r_0$ is an arbitrary spatial scale, $\phi$ is the azimuthal angle and the normalized axial coordinate is given by $\xi=z/(2k_0n_\infty r_0^2)$. In this representation the optical potential reads $V=2n_\infty k_0^2r_0^2\Delta n$. By expressing the mode $U=e^{i\mu \xi}e^{i\ell \phi}R(\eta)$ in terms of its orbital angular momentum $\ell$, and after using the radial transformation $R=\eta^{-1/2}u$ we reduce Eq.~\eqref{eq5} to a 1D form,
\begin{equation}
\label{eq6}
\left (-\frac{d^2}{d \eta^2}-V_{\textrm{eff}}(\eta)\right )u=-\mu u,
\end{equation}
with the effective potential $V_{\textrm{eff}}(\eta)=V(\eta)+\frac{1/4-\ell^2}{\eta^2}$. By designating the modes of Eq.~\eqref{eq6} as $u_{\ell m}$, having azimuthal and radial mode numbers $\ell$ and $m$ respectively, one can then generate an effective partner potential $V_{\textrm{eff}}^{(2)}(\eta)$ for a given effective potential $V_{\textrm{eff}}^{(1)}(\eta)$. As in the 1D case investigated before, these two potentials are related via the fundamental mode $u_{\ell_1 1}^{(1)}$ of the first potential; $V_{\textrm{eff}}^{(2)}=V_{\textrm{eff}}^{(1)}+2\frac{d^2}{d \eta^2}\ln \left (u_{\ell_1 1}^{(1)}\right)$. In the original coordinate system, $R_{\ell_1 1}^{(1)}=\eta^{-1/2}u_{\ell_1 1}^{(1)}$, which yields the following relation between the superpartner potentials:
\begin{equation}
\label{eq7}
V^{(2)}(\eta)=V^{(1)}(\eta)+2\frac{d^2}{d \eta^2}\ln \left (\eta^{\frac{\ell_1^2-\ell_2^2+1}{2}}R_{\ell_1 1}^{(1)}\right ).
\end{equation}
Note that in deriving the most general expression for $V^{(2)}$ we have assumed a different azimuthal mode number $\ell_2$ for the partner potential. In other words, a potential $V^{(1)}$ and its partner $V^{(2)}$, constructed for a certain $\ell_1$ and $\ell_2$, will only be supersymmetric with respect to the subsets $R_{\ell_1 m}^{(1)}$ and $R_{\ell_2 m}^{(2)}$ of their respective radial modes ($m=1,2,\ldots$). Note that the second term in Eq.~\eqref{eq7} may introduce a singularity at  $\eta=0$. Yet, this can be alleviated through an appropriate choice of $\ell_1$ and $\ell_2$. Near the origin ($\eta \ll 1$), the radial solutions $R_{\ell m}$ of any well-behaved potential $V(\eta)$ are proportional to $\eta^{|\ell|}$  \cite{suppl}, and thus $R_{\ell_1 1}(\eta)\sim \eta^{|\ell_1|}$. Therefore, Eq.~\eqref{eq7} yields a non-singular partner potential only if $|\ell_2|=|\ell_1|+1$. This relation reveals an unexpected result; in cylindrically symmetric settings, SUSY provides a link between sets of modes with adjacent azimuthal numbers. Given that  $V^{(1)}$ vanishes at $\eta \rightarrow \infty$ it then follows that \cite{suppl} $R_{\ell_1 1}^{(1)}\sim \frac{1}{\sqrt{\eta}}\exp\left ({-\sqrt{\mu_{\ell_1 1}}\eta}\right )$, and hence $V^{(2)}(\eta)\sim 1/\eta^2$ in this same limit.\\
\begin{figure}[t]
\begin{center}
\includegraphics[width=1\linewidth]{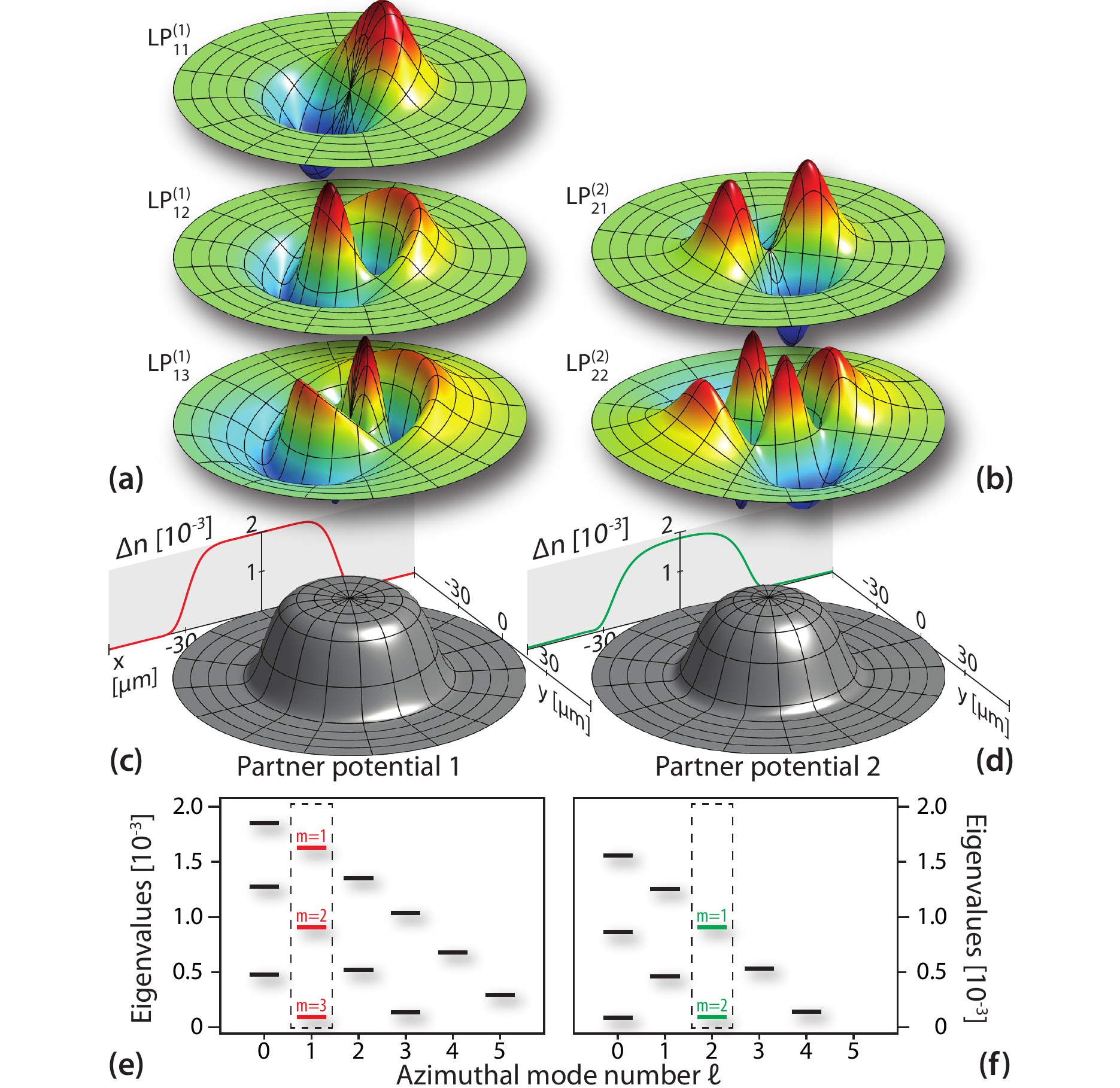}
\caption{(Color online). (a,b) Supersymmetric subsets of bound states corresponding to the SUSY pair of cylindrically symmetric index profiles (c,d) generated for azimuthal numbers $\ell_1=1$ and $\ell_2=2$. (e,f) Complete eigenvalue spectra (effective refractive indices) of both potentials. The respective subsets of SUSY states are indicated by dashed frames.}
\label{fig4}
\end{center}
\end{figure}
Figures~\ref{fig4}(a,b) depict the field profiles of the modes $LP_{\ell_{1,2} m}^{(1,2)}=e^{i\ell_{1,2}\phi}R_{\ell_{1,2} m}^{(1,2)}(\eta)$ corresponding to the two cylindrical superpartner index profiles in Figs.~\ref{fig4}(c,d). In this case, the original refractive index distribution is taken to be $\Delta n(r)=\delta e^{-(r/r_0)^8}$, where the core radius is $r_0=30\textrm{\textmu m}$, the index contrast amounts to $\delta =2\times 10^{-3}$ and the background refractive index is $n_\infty=1.5$. At a wavelength of $1.55\textrm{\textmu m}$, it supports a total of twelve guided  modes. Based on the lowest state with $\ell_1=1$, a partner potential for $\ell_2=2$ was generated according to Eq.~\eqref{eq7} [see Fig.~\ref{fig4}(d)]. Note that whereas SUSY holds between the modes with $\ell_1=1$ and $\ell_2=2$, the rest of the eigenvalues remain disjoint, as shown in Figs.~\ref{fig4}(e,f). By relating mode subsets of different azimuthal indices in this 2D setting, SUSY offers the possibility for a fully integrated realization of optical angular momentum multiplexing \cite{Wan2012}.\\
We next consider SUSY in finite periodic arrangements. For example, a lattice of $N$ well-separated single-mode waveguides is known to support a set of $N$ bound states or supermodes. In this array environment, the fundamental state is again node-free and hence can be readily used to generate a superpotential according to Eqs.~\eqref{eq2}. The corresponding SUSY partner resembles a lattice with $N-1$ dissimilar channels located in the gaps between the original waveguides (see~\cite{suppl}).\\
The coupled mode formalism provides an effective way to describe wave evolution in photonic lattices within the first band. The set of coupled differential equations \cite{Chr2003} for the modal field amplitudes $\mathbf{a}$ can be written in the form
$\mathcal{H}\mathbf{a}=\lambda \mathbf{a}$, where $\mathcal{H}$ is now the discrete Hamiltonian of the system. This discretization provides a powerful approach for constructing SUSY pairs: The Hamiltonian can be directly factorized using the Cholesky method \cite{Hog2006}. The pair of isospectral Hamiltonians thus obtained retains the tri-diagonal shape of $\mathcal{H}$, i.e. the SUSY partner represents again a photonic lattice with nearest-neighbor coupling. Note that whereas both Hamiltonians are $N\times N$ matrices, SUSY is
nevertheless unbroken in the sense that the $N^{\textrm{th}}$ waveguide of lattice 2 is completely decoupled.\\
\begin{figure}[b]
\begin{center}
\includegraphics[width=1\linewidth]{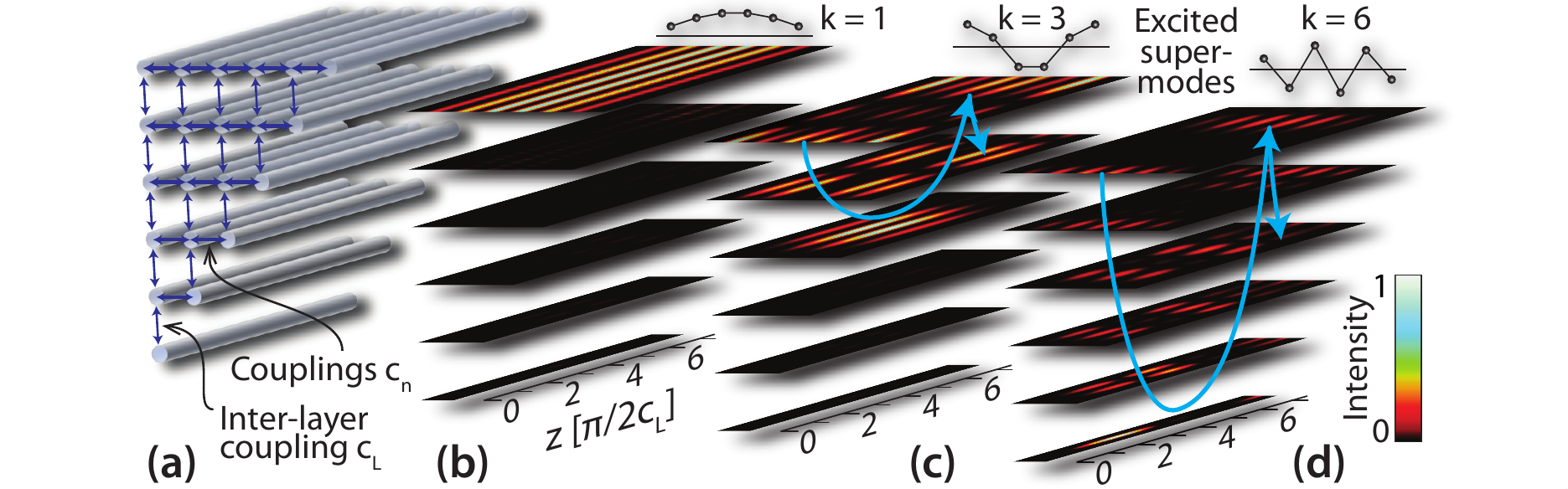}
\caption{(Color online). (a) Schematic of a SUSY ladder with $N=6$ layers. Propagation dynamics when a supermode of the original lattice is selectively excited. (b) $k=1$ (fundamental state): Confined in the first layer; (c) $k=3$: Penetrates only the first 3 layers (d) $k=6$: Moves freely across the entire ladder}
\label{fig5}
\end{center}
\end{figure}
Even more importantly, the discrete formalism outlined above relaxes the need for exactly controlling the refractive index landscape. In particular, the technological difficulties associated with sharp index depressions can be circumvented without any loss of functionality. Indeed, the control of only two parameters is here sufficient for the actual realization of SUSY optical systems: The waveguide's effective refractive index, which determines the propagation constant, and their separation, which relates to the coupling coefficient. A sequence of SUSY potentials can be iteratively obtained by discarding the respective isolated channels.  Such a SUSY ``ladder'' can facilitate a lossless decomposition of any input beam into its modal constituents. A weak coupling $c_L$ between such consecutive partner lattices, as indicated in Fig.~\ref{fig5}(a), does not perturb SUSY and allows for an interaction only between states with equal eigenvalues. Consequently, energy initially carried by the $k^{\textrm{th}}$ supermode in the fundamental lattice can be transported between all layers $1...k$, but is rejected by layer $k+1$. The propagation dynamics arising from the excitation of several supermodes in the fundamental lattice are shown in Figs.~\ref{fig5}(b-d) for such a SUSY ladder based on a uniform array with $N_0=6$ waveguides. The condition of weak inter-layer coupling was assured by setting $c_L$ to be $5\%$ of the coupling $C$ within the uniform lattice.\\
In conclusion we have shown that SUSY partner systems can be generated for any 1D refractive index landscape supporting at least one bound state. Despite their dissimilar shapes, SUSY structures can exhibit identical reflectivities and transmittivities for arbitrary angles of incidence. Subsequently the concept of optical SUSY was extended to 2D settings with cylindrical symmetry. In this case SUSY was established for sets of modes exhibiting consecutive azimuthal indices. In the context of photonic lattices, SUSY manifests itself as a reduction in the number of channels. This concept is general and highlights the potential of SUSY for robust optical filtering and signal processing applications.\\
We acknowledge financial support from NSF (grant ECCS-1128520) and AFOSR (grant FA95501210148). M.H. was supported by the German National Academy of Sciences Leopoldina (grant No. LPDS 2012-01).
\bibliography{GF}
\end{document}